\documentclass{elsart}
\usepackage{epsfig}
\usepackage{amssymb}
\usepackage{rotating}

\begin{document}

% Front matter, authorship and abstract %
%
\begin{frontmatter}

\title{Limits on spin-dependent WIMP-nucleon cross-sections
from the first ZEPLIN-II data}

{\center \large \bf The ZEPLIN-II Collaboration}

\author[RAL]{G. J. Alner},
\author[ICL,RAL]{H. M. Ara\'ujo},
\author[ICL]{A. Bewick},
\author[RAL,ICL]{C. Bungau},
\author[RAL]{B. Camanzi},
\author[SHE]{M. J. Carson},
\author[OXF]{R. J. Cashmore},
\author[SHE]{H. Chagani},
\author[LIP]{V. Chepel},
\author[UCL]{D. Cline},
\author[ICL]{D. Davidge},
\author[SHE]{J. C. Davies},
\author[SHE]{E. Daw},
\author[ICL]{J. Dawson},
\author[RAL]{T. Durkin},
\author[RAL,ICL]{B. Edwards},
\author[SHE]{T. Gamble},
\author[TEX]{J. Gao},
\author[EDI]{C. Ghag},
\author[ICL]{A. S. Howard},
\author[ICL]{W. G. Jones},
\author[ICL]{M. Joshi},
\author[EDI]{E. V. Korolkova},
\author[SHE]{V. A. Kudryavtsev\corauthref{cor1}},
\corauth[cor1]{Corresponding author; address: 
Department of Physics and Astronomy, University of Sheffield,
Sheffield S3 7RH, UK}
\ead{v.kudryavtsev@sheffield.ac.uk}
\author[SHE]{T. Lawson},
\author[ICL]{V. N. Lebedenko},
\author[RAL]{J. D. Lewin},
\author[SHE]{P. Lightfoot},
\author[LIP]{A. Lindote},
\author[ICL]{I. Liubarsky},
\author[LIP]{M. I. Lopes},
\author[RAL]{R. L\"{u}scher},
\author[SHE]{P. Majewski},
\author[SHE]{K Mavrokoridis},
\author[SHE]{J. E. McMillan},
\author[SHE]{B. Morgan},
\author[SHE]{D. Muna},
\author[EDI]{A. St.J. Murphy},
\author[LIP]{F. Neves},
\author[SHE]{G. G. Nicklin},
\author[UCL]{W. Ooi},
\author[SHE]{S. M. Paling},
\author[LIP]{J. Pinto da Cunha},
\author[EDI]{S. J. S. Plank},
\author[RAL]{R. M. Preece},
\author[ICL]{J. J. Quenby},
\author[SHE]{M. Robinson},
\author[TEX]{G. Salinas},
\author[UCL]{F. Sergiampietri}$^{,1}$,
\author[LIP]{C. Silva},
\author[LIP]{V. N. Solovov},
\author[RAL]{N. J. T. Smith},
\author[RAL,UCL]{P. F. Smith},
\author[SHE]{N. J. C. Spooner},
\author[ICL]{T. J. Sumner},
\author[ICL]{C. Thorne},
\author[SHE]{D. R. Tovey},
\author[SHE]{E. Tziaferi},
\author[ICL]{R. J. Walker},
\author[UCL]{H. Wang},
\author[TEX]{J. White} \&
\author[ROC]{F. L. H. Wolfs}

\address[RAL] {Particle Physics Department, STFC Rutherford Appleton 
Laboratory, UK}
\address[ICL] {Blackett Laboratory, Imperial College London, UK}
\address[SHE] {Department of Physics \& Astronomy, University of Sheffield, 
UK}
\address[OXF] {Brasenose College \& Department of Physics, 
University of Oxford, UK}
\address[LIP] {LIP--Coimbra \& Department of Physics of the 
University of Coimbra, Portugal}
\address[UCL] {Department of Physics \& Astronomy, University of 
California, Los Angeles, USA}
\address[EDI] {School of Physics, University of Edinburgh, UK}
\address[TEX] {Department of Physics, Texas A\&M University, USA}
\address[ROC] {Department of Physics \& Astronomy, University of 
Rochester, New York, USA}

\footnote{Also at INFN Pisa, Italy}

\newpage
\begin{abstract}
The first underground data run of the ZEPLIN-II experiment has set a limit
on the nuclear recoil rate in the two-phase xenon detector for direct dark matter
searches. In this paper the results from this run are converted into
the limits on spin-dependent WIMP-proton and WIMP-neutron cross-sections. 
The minimum
of the curve for WIMP-neutron cross-section
corresponds to $7 \times 10^{-2}$ pb at a WIMP mass of around 65~GeV.
\end{abstract}


\end{frontmatter}

% Main text %
%
% Introduction %
%
\section{Introduction}
Weakly Interacting Massive Particles (WIMPs) remain the most plausible
candidate for dark matter, responsible for about
80\% of the total matter and for about 23\% of
the total energy contents of the Universe.
WIMPs are expected to interact with ordinary matter via spin-independent
(sometimes also called scalar or coherent) and spin-dependent (axial)
interactions.
Spin-independent interactions of WIMPs should largely dominate in high-$A$
targets due to the $A^2$ coherence enhancement factor (here $A$ is
the atomic weight of the material used as a target). However, the relative
probability of the spin-independent and spin-dependent interactions depends
also on the particle content of WIMPs, i.e. on parameters of a particular
supersymmetric model. In some models, WIMPs interact predominantly
through the spin-dependent interactions. This stimulates the search for
spin-dependent interactions of WIMPs in addition to the spin-independent
case.

The interest in spin-dependent WIMP-nucleon interactions is enhanced by
the positive WIMP signal claimed by the DAMA Collaboration.
Although the first DAMA publications favoured spin-independent signal due
to the very soft spectrum of events observed, further studies allowed the
interpretation of the results in terms of the combination of both
spin-independent and spin-dependent interactions \cite{dama03}.

The spin-dependent WIMP-nucleus cross-section depends on the spin factor
of the nucleus that is primarily determined by the number
of protons and neutrons in the nucleus, namely whether it is odd or even.
For odd-proton nuclei the
spin-dependent WIMP-nucleus cross-section is mainly due to the
WIMP-proton interactions, whereas for odd-neutron nuclei it is dominated
by WIMP-neutron scattering.
For even-even nuclei the spin-dependent 
cross-section is highly suppressed.

Among the most sensitive targets for spin-dependent WIMP searches
are Na, I, Cs, F (all odd-proton) and Xe, Ge (having odd-neutron isotopes). 
The best limits from direct detection experiments
so far have been set by the NAIAD (WIMP-proton) \cite{naiad05} 
and CDMS \cite{cdms06},  ZEPLIN-I \cite{zeplin104} and EDELWEISS 
\cite{edelweiss05} (WIMP-neutron) experiments.

In this paper, the recent results of the ZEPLIN-II experiment are analysed
in terms of the spin-dependent limits. A brief description of the detector
and experimental data is given in Section 2. The method of calculating
spin-dependent limits is presented in Section 3 together with the results.
Conclusions are given in Section 4.

\section{ZEPLIN-II experiment}

ZEPLIN-II \cite{z2_a,z2_b} is a two-phase (liquid/gas) xenon detector 
searching for elastic scattering of WIMPs off xenon nuclei.
ZEPLIN-II is operated at the Boulby Underground Laboratory in the UK
at a depth of 2805~m~w.e. underground
with a muon flux of (4.09$\pm$0.15)$\times10^{-8}$ ~muons/cm$^2$/s
\cite{robinson03}.
The detector, data acquisition system, analysis procedure
and experimental data
are described in detail in Refs. \cite{smith07,carson07,luscher07}. We 
present here only a short summary of the detector performance and
some other features important for further analysis.

ZEPLIN-II consists of a vacuum cryostat containing
about 31~kg of liquid xenon. 
The target volume is viewed from above by seven 13~cm diameter 
photomultiplier tubes (PMTs).
The inner surface of the xenon vessel is covered by light-reflecting
PTFE for better collection of scintillation photons. 

The detector records
scintillation light and ionisation electrons from charged particles. 
Scintillation light is detected by the PMTs promptly
after excitation of the active medium. Ionisation electrons are drifted
through the liquid towards the surface by means of applied electric field, 
extracted from the
liquid into the gas where they are accelerated producing a secondary scintillation
signal as electroluminescence. 
The depth of liquid xenon (14~cm) corresponds to a maximum 
drift time of 73~$\mu$s for electrons at the drift field of 1~kV/cm.

To protect the target volume 
from radioactivity in rock (gammas and neutrons), the ZEPLIN-II detector is 
surrounded 
by hydrocarbon material and high-purity lead. Part of the hydrocarbon
shielding is instrumented liquid scintillator that also acts as an anticoincidence 
system (active veto) preventing the signals, detected simultaneously in the target
and in the veto, being interpreted as WIMP interactions. 

The trigger is provided by five-fold coincidences between different
PMTs at a single photoelectron level. Either a primary, S1, or a 
secondary, S2, signal can trigger the data acquisition. 
The signals from all seven PMTs
are recorded with
2~ns sampling time during 
$\pm 100$~$\mu$s around the trigger pulse, covering all
possible arrival times for both primary and secondary signals.
In an off-line data analysis three-fold coincidences between
different PMTs at a single photoelectron level were used to
identify and parameterise the primary signal S1.

A number of parameters have been measured for each waveform.
They are listed and discussed in Refs. \cite{smith07,carson07}, the most
important being the total areas of the primary and secondary pulses
(proportional to the number of photoelectrons),
S1 and S2, respectively, the time delay between them and the width
of the pulses (that determines whether the pulse is the primary or
secondary signal).

Position sensitivity of the experiment in the vertical direction
is achieved by considering
the time delay of the secondary pulse which is proportional to the
drift time of the electrons and thus determines the point of the interaction
along the drift field direction.
This allows us to exclude events originating on or close to the grid wires
that provide the electric field and are
contaminated mainly with radon progeny.
In the horizontal plane the event position is reconstructed using the relative
pulse areas from secondary signals on different PMTs. This method gives 
bigger uncertainty compared to the drift time due to the large PMT sizes and
small photon statistics at low energies.

A daily energy calibration of the detector using a
$^{57}$Co gamma-ray source allowed monitoring the stability of 
detector operation.
The WIMP/gamma discrimination performance of ZEPLIN-II has been tested
by calibrating the detector using high-energy gamma-ray ($^{60}$Co)
and neutron (AmBe) sources.
High-energy gamma-rays produce the main electron recoil background,
whereas fast neutrons scatter elastically off nuclei producing nuclear
recoils in the same way as expected from WIMP scatters.
Using the S2/S1 versus S1 plot from neutron calibration run
(Figure~8 in Ref. \cite{smith07}) the nuclear recoil acceptance box
has been defined as retaining 50\% of nuclear recoil events at any
given energy chosen for analysis. This acceptance box
was expected to have a small number of electron recoils due to the tail of
S2/S1 distribution observed in the gamma calibration run.

The first data run of the ZEPLIN-II detector had a live time of 31.2 days
after time periods with unstable operation conditions were removed
from the analysis. 
A number of software cuts have been applied to the
parameterised pulses allowing the selection of single interactions within
the fiducial volume of the detector to be made. These cuts are described
in Refs. \cite{smith07}. For each cut an energy dependent efficiency has been
evaluated either from the data or from a combination of data and simulations.

An important cut that reduces significantly the
fiducial volume of xenon, is the radial cut. This rejects the events that
are reconstructed as being close to the PTFE walls. In reality, due to
imperfect position reconstruction of the interaction points in the horizontal
plane, a long tail of events assumed to be originated at the walls is
reconstructed towards the centre of the detector. The majority of these events is 
believed to be caused
by the alpha decay of radon progeny accumulated on the charged surfaces.
To remove most of the `wall' events the radial cut
has been applied reducing the fiducial mass of xenon down to 7.2~kg
(see Ref. \cite{smith07} for more details).

Two xenon isotopes occurring naturally are sensitive to WIMP-nucleus
spin-dependent interactions: $^{129}$Xe and $^{131}$Xe. Other stable
xenon isotopes are even-even (with even numbers of proton and neutrons)
with very small coupling to WIMP spin. $^{129}$Xe and $^{131}$Xe are both
odd-neutron isotopes and hence are sensitive mainly to WIMP-neutron
interactions. Their relative abundances are 26.4\% and 21.2\%, respectively.
The effective exposure of these target isotopes to WIMPs is calculated as
$X_i = M \times t \times C_i \times A \times \epsilon$, where $X_i$ is the exposure 
of the {\it i}th
isotope, $M=7.2$~kg is the fiducial mass of the target, $t=31.2$~days is the live time
of the run, $C_i$ is the relative abundance of the isotope, $A=0.5$ is the
fraction of nuclear recoils in the acceptance window on S2/S1 vs S1 plane
chosen for the analysis and $\epsilon$ is the overall energy dependent efficiency 
of other
cuts described above and in Ref. \cite{smith07}. 
If we neglect the energy dependent efficiency $\epsilon$
which increases from about 35\% at 5~keV to about 75\% at 10 keV, the
exposures for different xenon isotopes are 29.7~kg$\times$days for $^{129}$Xe 
and 23.8~kg$\times$days for $^{131}$Xe. Cut efficiency $\epsilon$
further reduces the exposure making it also energy dependent.

The energy range of 5-20~keV (electron equivalent) 
has been chosen for the data analysis.
Below 5~keV the trigger efficiency is rather small (less than 40\%).
Above 20~keV the sensitivity of xenon target to WIMP interactions
decreases significantly because of the rapidly falling form-factor.
To avoid any bias, the selection of cuts and energy range for the final analysis 
was based on the results from calibration runs and from `unblinded' 10\% of
data prior to 'opening the box' with the remaining 90\% of data.

The analysis of data has revealed 29 events
in the nuclear recoil acceptance box, the expected background rate
due to electron recoils and `wall' events being $28.6\pm4.3$ (see \cite{smith07}
for full description of the procedure to evaluate the expected background).
Based on the previously published simulations 
\cite{smith04,carson04,araujo05b,bungau05}
we expect to have less than 1 nuclear recoil from neutron background
in the detector for the aforementioned exposure. 

Applying the procedure described by Feldman and Cousins \cite{feldman98}
the 90\% CL upper limit on the number of nuclear recoils has been set as 
10.4 using the ROOT software \cite{root} and then 
converted into an upper
limit on the WIMP-nucleon spin-independent cross-section \cite{smith07}.
The limit on the nuclear recoil rate can also be used to set a limit
on spin-dependent interactions. This procedure is described below.

\section{Limits on spin-dependent cross-sections}

In the spin-independent 
case Majorana WIMP coupling to protons and neutrons is very similar,
the coherence enhancement factor is proportional to $A^2$ and the 
cross-section on the nucleus does not depend on particular WIMP model
parameters.
For spin-dependent interactions the coupling to protons and neutrons is
very much different and depends strongly on the WIMP model parameters.
In the derivation of the spin-dependent limits we follow the procedure 
described in Ref. \cite{tovey00}. 

The WIMP-nucleus cross-section, $\sigma_A$, can be written as:

\begin{equation}
\sigma_A = \frac{32}{\pi} G_F^2 \mu_A^2 (a_p <S_p> + a_n <S_n>)^2 \frac{J+1}{J},
\label{sigmaA}
\end{equation}

where $G_F$ is the Fermi weak coupling constant, $\mu_A$ is the WIMP-nucleus 
reduced mass, $a_{p,n}$ are the effective WIMP-proton (neutron) couplings, 
$<S_{p,n}>$ are the expectation values of the proton and neutron spins in the 
nucleus (or spin-factors) and $J$ is the nuclear spin. For a proton or a neutron as a target, 
Eq. (\ref{sigmaA}) is transformed into the cross-section for WIMP-proton (neutron)
interactions with the proton (neutron) spins $<S_{p,n}> = 1/2$ and $J=1/2$.
Eq. (\ref{sigmaA}) does not correspond to the total WIMP-nucleus
cross-section, but to that at zero momentum transfer. As in the case for
spin-independent interactions, the cross-section for zero momentum transfer
is usually presented for comparison with other results and model predictions.
To evaluate the `real' interaction cross-section and to compare it with the experimental 
data this has to be multiplied by
the nuclear form-factor $F^2(q)$ which is a function of the momentum
transfer, $q$ \cite{lewin96}. 

 As in Ref. \cite{tovey00} we assume that the total WIMP-nucleus cross-section
 at zero momentum transfer is dominated by either WIMP-proton or
 WIMP-neutron interactions only, setting the 2nd component equal to 0.
 In this case the WIMP model dependent parameters are cancelled in the
 equation for the ratio of the cross-sections:
 
\begin{equation}
\frac{\sigma_{p,n}}{\sigma_A} = \frac{3}{4} \frac{\mu_{p,n}^2}{\mu_A^2} 
\frac{1}{<S_{p,n}>^2} \frac{J}{J+1},
\label{cr-ratio}
\end{equation}

which retains only the nuclear physics parameters. As $\sigma_A$ is measured
in an experiment or a limit on $\sigma_A$ is set from the experimental data 
(assuming a particular form-factor) and nuclear physics parameters 
$<S_{p,n}>$ can be
calculated independently of the WIMP model, the WIMP-proton (neutron)
cross-section can be evaluated in a (almost) model-independent way
using Eq.~(\ref{cr-ratio}).
The only WIMP model dependence that remains is hidden in the form
factor that appears to be different for different WIMP models.
This dependence, however, is not significant for most isotopes 
and is comparable to the
uncertainty in the nuclear physics model used.

For xenon as a target, additional complication comes from the existence of 
two isotopes with odd-neutron nuclei, $^{129}$Xe and $^{131}$Xe. 
In this case
the combined limit on the cross-section is calculated as \cite{lewin96}:

\begin{equation}
\frac{1}{\sigma_{p,n}} = \sum_{A_i} \frac{1}{\sigma_{p,n}^{A_i}},
\label{combined}
\end{equation}

where $\sigma_{p,n}^{A_i}$ are the WIMP-proton (neutron)
limits set for cross-sections on 
individual isotopes taking into account their fraction by weight, and
$\sigma_{p,n}$ is the combined limit on the WIMP-proton (neutron)
cross-section.

In further calculations we used the spherical isothermal dark matter 
halo model with the following parameters:
$\rho_{dm}$ = 0.3 GeV cm$^{-3}$, $v_o$ = 220 km/s, 
$v_{esc}$ = 600 km/s and $v_{Earth}$ = 232 km/s.
The form-factors were computed
for the two xenon isotopes using the 
nuclear shell model calculations \cite{ressell97}
with `Bonn A' nucleon-nucleon potential. This is based on a comprehensive
meson-exchange model for the nucleon-nucleon interaction in field theory
developed by the Bonn group \cite{machleidt87}.
This model offers the most consistent approach to the
nuclear many-body problem at low energies relevant to WIMP interactions.
It includes all important diagrams with a total exchanged mass up to 
about the cutoff mass ($\sim1$~GeV).
The various meson-exchange contributions in this mass range 
are introduced step by step
proceeding from lowest-order to higher-order processes and from long 
range to short range. The model predictions agree well with the deuteron data
and parameters derived from nucleon-nucleon scattering experiments
\cite{machleidt87}.

The total form-factor can be written in the form:
\begin{equation}
F^2(q) = \frac{S(q)}{S(0)},
\label{formfactor}
\end{equation}

where 

\begin{equation}
S(q)  = a_0^2 S_{00} (q) + a_1^2 S_{11} (q) + a_0 a_1 S_{01} (q).
\label{structure}
\end{equation}

Here $a_0=a_p+a_n$, $a_1=a_p-a_n$ and $S_{00}$, $S_{11}$,
$S_{01}$ are the isoscalar, isovector and interference contributions to the
spin structure function $S(q)$, respectively.
These contributions are determined by a nuclear model (independent of
the WIMP type), with coefficients $a_0$ and $a_1$
related to the WIMP-proton and WIMP-neutron coupling constants $a_{p,n}$ 
(and hence WIMP-type dependent). The coupling coefficients
depend also on the assumption about the quark spin fractions, i.e.
the fractional contributions of different quark species to the nucleon spin.
We used the WIMP-proton and WIMP-neutron coupling constants
$a_p$ and $a_n$ with the quark spin fractions from Ref. \cite{ellis95}.
Although the form-factor is normalised to the value at zero momentum
transfer (Eq. (\ref{formfactor})) that also includes the WIMP coupling coefficients, 
they do not cancel out
completely, leaving some dependence on the WIMP particle model.
The dependence on the WIMP model parameters for xenon isotopes
is not as weak as,
for instance, for iodine and low-A isotopes. The form-factor
for higgsino interactions, however, is the smallest among
WIMP particle models, so the limit based on the present calculations with 
higgsino form-factor is conservative.

The spin-factors $<S_{p,n}>$ for the two odd-neutron Xe isotopes
have been calculated in Ref. \cite{ressell97} and also given
in Ref. \cite{tovey00}. In the present analysis we used the values 
reported for the same nuclear model with Bonn A potential as
for the form-factors.

\begin{figure}[htb]
   \centerline{\epsfig{file=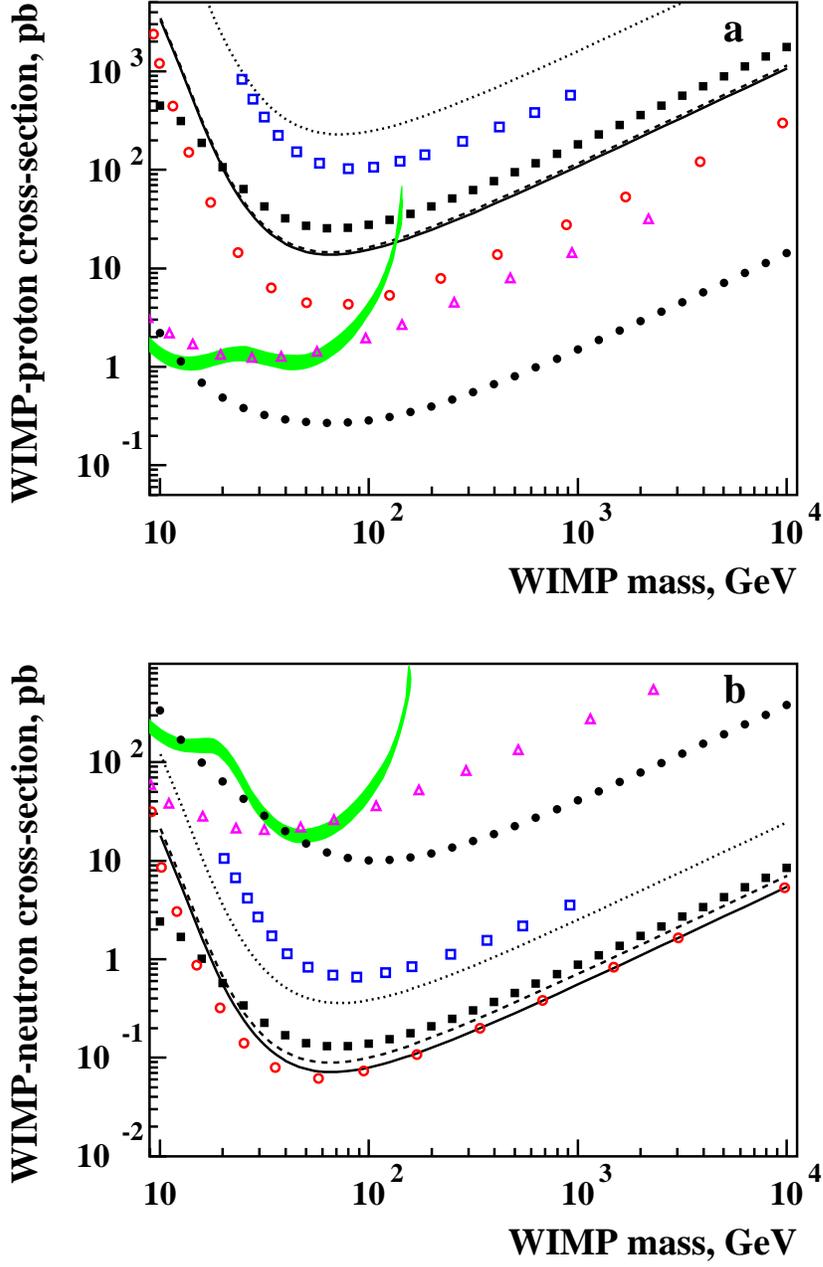,width=12.cm}}
    \caption{90\% CL upper limits on the WIMP-proton (a) and WIMP-neutron (b)
    spin-dependent cross-sections derived from ZEPLIN-II data. 
    The limits from the two isotopes ($^{129}$Xe -- dashed curves, 
    $^{131}$Xe -- dotted curves) and the combined
    limit (solid curves) are presented. The limits from some other
    experiments are also shown: NAIAD \cite{naiad05} ($\bullet$), ZEPLIN-I
    \cite{zeplin104} ($\blacksquare$), CDMS \cite{cdms06} ($\circ$), EDELWEISS
    \cite{edelweiss05} ($\square$) and PICASSO \cite{picasso05} ($\bigtriangleup$)
    (the latter result coincides with that of the SIMPLE experiment \cite{simple05}).
    The interpretation of the positive annual modulation signal observed
    by the DAMA experiment \cite{dama03}, in terms of the constraints on
    spin-dependent cross-sections reported by Savage et al. \cite{savage04},
    is shown by the filled area.}
  \label{limits}
\end{figure}

Figure \ref{limits} shows the 90\% CL limits on the 
WIMP-proton (a) and
WIMP-neutron (b) spin-dependent cross-sections calculated using the 
ZEPLIN-II data and the procedure described above.
Results from some other experiments are also shown.
The ZEPLIN-II limits are also given in numerical form in Table \ref{tlimits}
to allow more accurate comparison with other experiments.
The minimum
of the curve for the WIMP-neutron cross-section
corresponds to $7 \times 10^{-2}$ pb. The limits
are dominated by the contribution from the $^{129}$Xe isotope. The 
ZEPLIN-II limits on WIMP-neutron cross-section are comparable
to the currently best result obtained by the CDMS experiment \cite{cdms06}.
Although the CDMS spin-independent limits are better than
those from ZEPLIN-II, the spin-dependent WIMP-neutron cross-section limits 
presented here are very
similar to those set by CDMS due to the higher fraction of odd-nucleon
isotopes in xenon.

The uncertainties in the nuclear spin and form-factors used for the evaluation 
of limit are not negligible. 
Apart from the remaining dependence on the particle model, there is an 
uncertainty related to the nuclear model.
Ressell and Dean \cite{ressell97} found a factor of 2
difference between their calculations of the spin structure functions for $^{131}$Xe 
(higher values at zero momentum transfer) and earlier
calculations by Engel \cite{engel91}. The model used by Ressell and 
Dean \cite{ressell97} with Bonn A potential 
gives smaller (again by a factor of 2 approximately) 
spin structure functions than simple 'single particle' model. 
Slightly lower spin expectation
value $S_n$ (by about 20\%) for $^{129}$Xe was obtained with Nijmegen II 
potential \cite{ressell97} which results in a 20\% higher limit on the
cross-section. This makes the systematic uncertainty of the cross-section
limit as large as a factor of 2 due to the nuclear model calculations.
We stress, however, that the model used in the present analysis is based
on the most recent and accurate calculations of nuclear spin and form-factors
\cite{ressell97}.

The predictions of different nuclear models for different nuclei are not strongly
correlated in the sense that two models can predict, for instance, similar
spin expectation values for one nucleus but very much different spins for another
nucleus. Thus, possible future detection of WIMP interactions with several 
different target nuclei should help in reducing the uncertainties in nuclear 
physics models and, hence, in improving accuracy of the WIMP parameters'
estimates.

The limits shown in Figure~\ref{limits} were obtained for a pure higgsino as
a WIMP.
Assuming pure photino or bino as a WIMP gives 35-40\% better
limit on the cross-section, whereas the limit for pure zino would be a factor
of 3 lower. As precise composition of WIMPs is not known, the assumption
of a pure higgsino leads to the most conservative limit.

To assess the significance of the spin-dependent limits less than
0.1~pb, it should be noted that a marginal dark matter candidate,
the heavy Majorana neutrino, could have a spin-dependent cross-section
of about 0.01~pb. However, a re-evaluation of likely supersymmetric
candidates by Ellis \cite{ellis2000} suggests spin-dependent
cross-sections no greater than $10^{-4}$~pb.

\begin{figure}[htb]
   \centerline{\epsfig{file=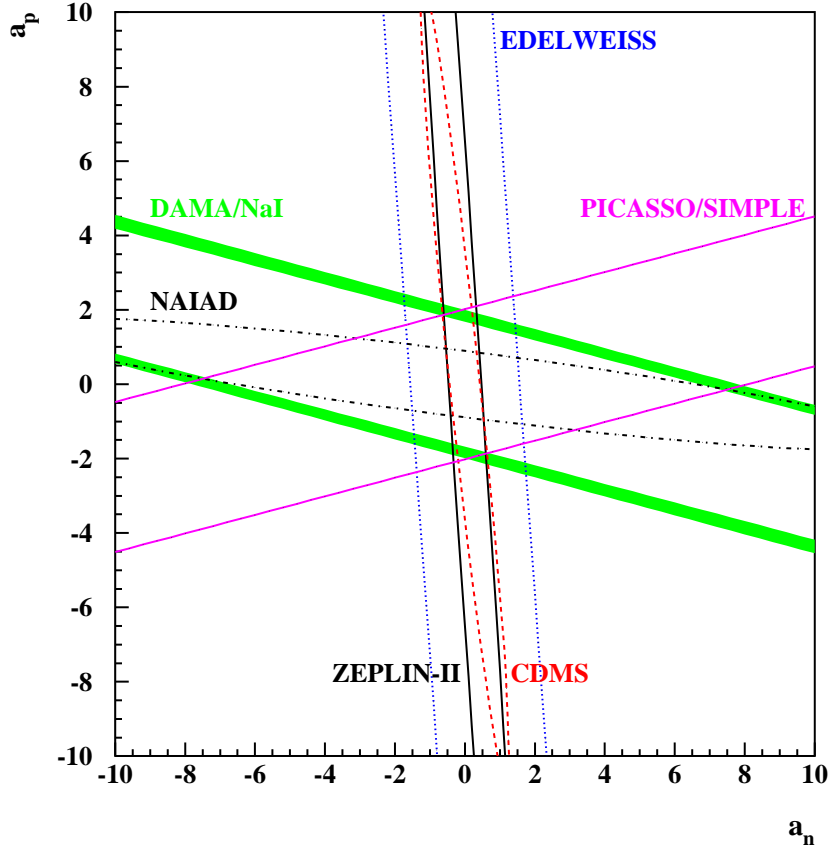,width=12.cm}}
    \caption{Constraints on the WIMP-proton and WIMP-neutron
    coupling coefficients $a_p$ and $a_n$ (for a WIMP mass of 50~GeV)
    as derived from
    90\% CL upper limits on the WIMP-proton and WIMP-neutron 
    cross-sections from different experiments. 
    The ZEPLIN-II results are shown by solid curves (almost vertical parallel lines).
    The region between the two lines is allowed by the ZEPLIN-II data.
    Other results are: CDMS (dashed curves) \cite{cdms06}, EDELWEISS
    (dotted curves) \cite{edelweiss05}, NAIAD (dashed-dotted curves, nearly
    horizontal) \cite{naiad05} and PICASSO/SIMPLE (solid curves, nearly 
    horizontal) \cite{picasso05,simple05}. The allowed regions from the DAMA/NaI
    experiment \cite{dama03} is shown by the filled area as calculated from the
    cross-section allowed regions reported in Ref.~\cite{savage04}.}
  \label{alimits}
\end{figure}

Figure~\ref{alimits} shows constraints on the WIMP-proton and WIMP-neutron
coupling coefficients $a_p$ and $a_n$ (for a WIMP mass of 50~GeV)
as derived from the ZEPLIN-II data in comparison with other experiments. 
Here again we used the formalism described in Ref.~\cite{tovey00} that allows
conversion of the cross-section limits into the allowed regions on the
$a_p-a_n$ plane using the equation:

\begin{equation}
\sum_{A_i} \left ( \frac{a_p}{\sqrt{\sigma_p^{A_i}}} \pm  \frac{a_n}{\sqrt{\sigma_n^{A_i}}} 
\right ) ^2 < \frac{\pi}{24 G_F^2 \mu_p^2},
\label{eq-alimits}
\end{equation}

where the small mass difference between the proton and the neutron is ignored.

Strictly speaking the procedure is mathematically correct only if the form-factor
used to evaluate the limits on cross-sections $\sigma_{p,n}^{A_i}$
is independent of the WIMP model. In most cases (including ZEPLIN-II
limit derivation in this paper) the form-factor depends on the WIMP model,
i.e. on the WIMP-proton and WIMP-neutron coupling coefficients
$a_{p,n}$. In practice this means that the limits $\sigma_{p,n}^{A_i}$ depend
on $a_{p,n}$. We used the form-factor for pure higgsino and in the derivation
of the allowed regions on the $a_p - a_n$ plane we ignored the dependence
of the form-factor, defined as $F^2(q) = S(q) / S(0)$, on the coupling coefficients.
This remark concerns also other experiments 
\cite{naiad05,edelweiss05,picasso05,simple05}
for which similar procedure was used. We repeated the derivation of the
allowed regions on the $a_p - a_n$ plane from the cross-section limits
from these experiments and found them to be in good agreement with
the original publications \cite{edelweiss05,picasso05,simple05}.
The allowed region from the CDMS experiment was copied from
Ref. \cite{cdms06}. Our computation of the allowed region for CDMS
gave slightly different result which implied that the CDMS Collaboration used another
procedure for constraining coupling coefficients in Ref.~\cite{cdms06}. 
The allowed region from the DAMA/NaI
experiment \cite{dama03} was calculated using the allowed regions for the
cross-sections from Ref.~\cite{savage04}.

One of the alternative
methods of setting limits on the coupling constants was suggested in 
Ref.~\cite{savage04}. Despite its complexity,
this method has an advantage of providing model independent 
constraints on the coupling coefficients by using the spin structure
function $S(q)$ directly in the process of the data analysis, i.e. the coupling
coefficients $a_{p,n}$ being free parameters in the fit to the data.
Note, however, that for accurate comparison between different
experiments, all data have to be analysed using the same method.

Figure~\ref{alimits} shows that for 50~GeV WIMP mass the DAMA allowed
region (filled area on the figure) is excluded by the combination of other
experiments. Savage et al.~\cite{savage04}, however, found that
the interpretation of the DAMA positive signal in terms of spin-dependent
interactions is still compatible with other experiments at small WIMP
masses (5--13~GeV). NAIAD (sensitive mainly to $a_p$) and 
CDMS/ZEPLIN-II (sensitive mainly to $a_n$) provided so far the most
stringent constraints on the WIMP-nucleon spin-dependent interactions.

\section{Conclusions}
The upper limits on the WIMP-proton and WIMP-neutron
spin-dependent cross-sections have been set using the ZEPLIN-II
data. The minimum
of the curve for WIMP-neutron cross-section
corresponds to $7 \times 10^{-2}$ pb. 
The limits on WIMP-neutron cross-section are comparable
to the currently best result obtained by the CDMS experiment.

\section{Acknowledgements}
This work has been funded by the UK Particle Physics And Astronomy 
Research Council (PPARC), the US Department of Energy (grant numbers
DE-FG03-91ER40662 and DE-FG03-95ER40917) and the 
US National Science Foundation (grant number 
PHY-0139065).  We acknowledge support from the Central Laboratories for 
the Research Councils (CCLRC), the Engineering and Physical Sciences 
Research Council (EPSRC), the ILIAS integrating activity 
(Contract R113-CT-2004-506222), the INTAS programme 
(grant number 04-78-6744) and the 
Research Corporation (grant number RA0350). We also acknowledge 
support from Funda\c{c}\~ao para a Ci\^encia e Tecnologia 
(project POCI/FP/FNU/63446/2005), the Marie Curie International 
Reintegration Grant (grant number 
FP6-006651) and a PPARC PIPSS award (grant PP/D000742/1). We would 
like to gratefully acknowledge the strong support of Cleveland Potash Ltd., the 
owners of the Boulby mine, and J. Mulholland and L. Yeoman, the 
underground facility staff. 
We would like to thank the team from ITEP, Moscow, led by Dr. D. Akimov
for their valuable contribution to the ZEPLIN programme.
We acknowledge valuable advice from Prof. R. 
Cousins on the application and extension of the Feldman-Cousins tables.

\pagebreak

\begin{table}[htb]
\caption{90\% CL limits on spin-dependent WIMP-neutron, 
$\sigma_n$, and WIMP-proton, $\sigma_p$,
cross-sections as functions of WIMP mass, $M_{W}$, 
for $^{129}$Xe and $^{131}$Xe isotopes together with the
combined limits from the ZEPLIN-II experiment.}
\vspace{1cm}
\begin{center}
\begin{scriptsize}
\begin{tabular}{|c|c|c|c|c|c|c|}\hline
$M_{W}$ (GeV) & $\sigma_n$ (pb),  $^{129}$Xe & $\sigma_n$ (pb), $^{131}$Xe & 
$\sigma_n$ (pb) & $\sigma_p$ (pb),  $^{129}$Xe & $\sigma_p$ (pb), $^{131}$Xe & 
$\sigma_p$ (pb)\\
\hline
10 &  $2.11\times10^{1}$ & $1.21\times10^{2}$ & $1.79\times10^{1}$ &
 $3.46\times10^{3}$ & $7.71\times10^{4}$ & $3.31\times10^{3}$ \\
20 & $6.51\times10^{-1}$ & $3.47\times10^{0}$ & $5.48\times10^{-1}$ &
$1.07\times10^{2}$ & $2.21\times10^{3}$ & $1.02\times10^{2}$ \\
40 & $1.14\times10^{-1}$ & $5.14\times10^{-1}$ & $9.31\times10^{-2}$ &
$1.87\times10^{1}$ & $3.28\times10^{2}$ & $1.77\times10^{2}$ \\
63 & $8.89\times10^{-2}$ & $3.64\times10^{-1}$ & $7.15\times10^{-2}$ &
$1.46\times10^{1}$ & $2.33\times10^{2}$ & $1.38\times10^{1}$ \\
100 & $1.00\times10^{-1}$ & $3.84\times10^{-1}$ & $7.94\times10^{-2}$ &
$1.64\times10^{1}$ & $2.45\times10^{2}$ & $1.54\times10^{1}$ \\
200 & $1.61\times10^{-1}$  & $5.87\times10^{-1}$ & $1.26\times10^{-1}$ &
$2.64\times10^{1}$ & $3.75\times10^{2}$ & $2.47\times10^{1}$ \\
400 & $2.96\times10^{-1}$ & $1.06\times10^{0}$ & $2.31\times10^{-1}$ &
$4.86\times10^{1}$ & $6.76\times10^{2}$ & $4.54\times10^{1}$ \\
630 & $4.57\times10^{-1}$ & $1.63\times10^{0}$ & $3.57\times10^{-1}$ &
$7.51\times10^{1}$  & $1.04\times10^{3}$ & $7.01\times10^{1}$ \\ 
1000 & $7.14\times10^{-1}$ & $2.53\times10^{0}$ & $5.57\times10^{-1}$ &
$1.17\times10^{2}$ & $1.62\times10^{3}$ & $1.09\times10^{2}$ \\
10000 & $6.99\times10^{0}$ & $2.46\times10^{1}$ & $5.45\times10^{0}$ &
$1.15\times10^{3}$ & $1.57\times10^{4}$ & $1.07\times10^{3}$ \\
\hline
\end{tabular}
\end{scriptsize}
\end{center}
\label{tlimits}
\end{table}

\pagebreak

\end{document}